\documentclass[
reprint,
superscriptaddress,
aps,prb
]{revtex4-2}

\usepackage[british]{babel} 
\usepackage[autostyle]{csquotes}
\usepackage{amsmath}
\usepackage{amssymb}
\usepackage{graphicx}
\usepackage{subcaption}
\usepackage{xcolor}
\usepackage{upgreek}

\newcommand{\sslash}{\mathbin{/\mkern-6mu/}} 

\begin{document}
	\title{Efficient computation of thermal radiation from biperiodic layered systems using the T-matrix method}
	\author{Martin Gabbert}
	\affiliation{Institute of Theoretical Solid State Physics, Karlsruhe Institute of Technology, Kaiserstr. 12, 76131 Karlsruhe, Germany}
	\author{Markus Nyman}
	\affiliation{Institute of Nanotechnology, Karlsruhe Institute of Technology, Kaiserstr. 12, 76131 Karlsruhe, Germany}
	\author{Lukas Rebholz}
	\affiliation{Institute of Theoretical Solid State Physics, Karlsruhe Institute of Technology, Kaiserstr. 12, 76131 Karlsruhe, Germany}
	\author{Carsten Rockstuhl}
	\affiliation{Institute of Theoretical Solid State Physics, Karlsruhe Institute of Technology, Kaiserstr. 12, 76131 Karlsruhe, Germany}
	\affiliation{Institute of Nanotechnology, Karlsruhe Institute of Technology, Kaiserstr. 12, 76131 Karlsruhe, Germany}
	\author{Ivan Fernandez-Corbaton}
	\affiliation{Institute of Nanotechnology, Karlsruhe Institute of Technology, Kaiserstr. 12, 76131 Karlsruhe, Germany}
	
	\begin{abstract}
	Metasurfaces are becoming important tools for the control of thermal radiation. Understanding their functional possibilities on computational grounds requires evaluating the response of the biperiodic layered system for many degrees of freedom, including several radiation directions and polarisations, while varying lattice spacing, thicknesses, and/or materials of homogeneous layers, over a range of frequencies. 
	The diverse set of cases that need to be considered in simulations prompts for efficient numerical tools to handle them. To respond to this need, we present a method for computing the thermal radiation from metasurfaces that combines the directional Kirchhoff law with efficient T-matrix based calculations. We show that such a method can accurately reproduce experimental data from a metasurface made of platinum square plates. 
	Additionally, we predict highly circularly polarised emissivity from a chiral metasurface. When comparing CPU-times, the method outperforms other approaches such as rigorous coupled wave analysis already at the modest number of 61 cases per frequency. 
	\end{abstract}
	
	\maketitle\newpage
	
	\section{Introduction}
	The recent developments in nanotechnology have enabled many new possibilities for constructing two-dimensional periodic arrays of nanoparticles, known as metasurfaces, which enable a refined control of electromagnetic fields. In particular, the design and fabrication of metasurfaces for the control of thermal radiation is an active area of research \cite{greffet-2002,eriksen-2024,costantini-2015,blanchard-2021,fernandez-hurtado-2017,overvig-2021,perrakis-2020}, with timely applications such as thermophotovoltaic devices and radiative cooling \cite{song-2015,cuevas-2018,picardi-2023}. Numerical methods are used to predict and optimise the desired properties of the metasurfaces \cite{elsawy-2020,schulz-2024,gallinet-2015} before their fabrication and experimental characterisation.

    A key trade-off in these numerical methods is between accuracy and computational cost. 
	While being computationally costly, finite-element methods (FEMs) have become a standard for retrieving reliable reference data because of their high accuracy \cite{monk-2003}. On the other hand, other methods such as the rigorous coupled wave analysis (RCWA), also known as the Fourier modal method \cite{li-1997}, are applied for their faster computation time while remaining reasonably accurate in suitable planar systems \cite{moharam-1995}. 

    This work introduces a method for computing angle- and polarisation-resolved thermal emission spectra of metasurfaces. It uses a set of pre-computed T-matrices, also known as transition matrices \cite{waterman-1965}, and the publicly available \textsc{treams} software library \cite{beutel-2023-treams} for constructing metasurfaces from the T-matrices of their unit cells, and computing their optical properties. 
	The thermal spectra are then computed using the directional Kirchhoff law of thermal emission \cite{kirchhoff-1860}, together with the common assumption that the thermal emission of an object at temperature $T$ is independent of its environment \cite[\S~2]{baltes-1976}.
	In this work, the T-matrices have been computed using the FEM. 
	We show that the proposed method is able to reproduce experimental data \cite{costantini-2015} and make predictions about more exotic metasurfaces that have not yet been fabricated. 
	Furthermore, the proposed method can compete with more established approaches such as pure FEM or RCWA in terms of computational costs. 
	Once the T-matrix has been obtained, all frequency-independent computations can be performed within \textsc{treams} in a small fraction of the time that the other solvers require. 
	Therefore, computing optical properties via the T-matrix is an alternative to pure FEMs or RCWA and allows for efficient computation of angle- and polarisation-resolved spectra because a single T-matrix contains all optical properties for a given frequency. 
	We estimate that the method will outperform even RCWA for most metasurfaces. Importantly, the number of cases per frequency, and hence the computational advantage of the proposed method will quickly increase as more degrees of freedom are exploited when optimising the design. 
	By `cases', we denote all combinations of degrees of freedom of the metasurface that do not change the meta-atoms of which the metasurface is composed. Examples of such degrees of freedom are the lattice kind, lattice spacing, and the material and thickness of homogeneous layers embedding the lattice. 

    At this point, it is worth mentioning a current limitation of the T-matrix method, namely, the difficulties that arise when computing the interactions between two objects that penetrate each other's smallest circumscribed spheres. This would affect planar periodic metasurfaces made from tall and thin objects located very closed to each other. This issue is under investigation, and there are some proposed solutions \cite{theobald-2017,egel-2017-CUDA,martin-2019,schebarchov-2019,rother-2021,barkhan-2021,lamprianidis-2022}.

    The rest of the article is organised as follows. 
	Section \ref{sec:theory+methods} outlines the theory and methods used in the implementation. In Section \ref{sec:Pt_slabs}, the method is applied to a metasurface of platinum particles, and shown to satisfactorily reproduce experimental measurements \cite{costantini-2015}. 
	Further validation of the method is provided by comparison with RCWA simulations. The computational costs are discussed, including a third option using a full FEM code. 
	Section \ref{sec:Ag_helices} contains the analysis of the thermal emissivity of a chiral metasurface, which consists of a square lattice of copies of a silver helix which was optimised for maximising its electromagnetic chirality (em-chirality) \cite{fernandezcorbaton-2016} at a particular frequency. The chiral metasurface features two frequency regions that simultaneously show significant emissivity and a large degree of circular polarisation, each region being dominated by a different handedness. Finally, Section \ref{sec:conclusion} concludes the article. 

    In the two examples contained in this article, the sizes of the objects in the lattices are of the order of one to a few micrometers. At such scales, the T-matrices are obtained using methods that solve the Maxwell equations. The methods can be generic such as FEM, finite-difference time-domain, or boundary element methods \cite{hohenester-2022}, or specific for particular geometries \cite{doicu-1999,mackowski-2002}. 
	The T-matrix, however, is a rather generic model which can also be obtained for much smaller objects, such as molecules, using quantum-chemical methods \cite{fernandezcorbaton-2020,zerulla-2022-moleculecavity}. This allows one to compute the thermal radiation of BINOL molecules \cite{mazo-vasquez-2025}, for example. 
	Therefore, the method presented here can be applied beyond arrays of micro-particles, to less common systems containing layers of molecular crystals \cite{van-lam-2024}, or graphene patches \cite{eriksen-2024}. 

    We foresee that the presented combination between the popular T-matrix formalism \cite{wriedt-2007,hellmers-2009,gouesbet-2024,mishchenko-2019}, and the efficient treatment of lattices in \textsc{treams} will become a valuable tool for the design of metasurfaces for applications in thermal photonics. {\color{blue}**T-matrices and python code to be made available with links indicated here**}

    \section{Theory and methods}\label{sec:theory+methods}
    First, to describe the optical response of an isolated scatterer of a metasurface, we made use of the T-matrix formalism that relates incident and scattered field by a linear operator, the T-matrix. This operator fully describes the linear light-matter interaction between light and a given object. The T-matrix is typically expressed in the basis of multipolar fields \cite{waterman-1965}. 
	There are different methods for computing T-matrices, for example the extended boundary condition method \cite{mishchenko-1996}, the FEM \cite{fruhnert-2017}, and the discrete dipole approximation method \cite{mackowski-2002}. 
	
	To compute the thermal radiation from a given metasurface, the emissivity $\mathcal{E}(\hat{r},\tau,\nu)$ in some direction $\hat{r}$ is determined from the absorptivity $\mathcal{A}(\hat{r},\tau,\nu)$, according to Kirchhoff's law of thermal radiation \cite{kirchhoff-1860}, 
	\begin{equation}\label{eq:Kirchhoff}
		\mathcal{A}(\hat{r},\tau,\nu) = \mathcal{E}(-\hat{r},\tau,\nu). 
	\end{equation}
    The emissivity denotes the ratio of radiation intensity emitted by the metasurface to the radiation intensity that would be emitted by a black body at the same temperature $T$, at frequency $\nu$ with polarisation $\tau = \pm1$, in direction $\hat{r}$. 
    In spherical coordinates, the transformation $\hat{r}\to-\hat{r}$ to the opposite direction is equivalent to $(\theta,\varphi)\to(\pi - \theta, \varphi + \pi)$. 
	The meaning of the polarisation index $\tau$ depends on which polarisation basis (circular or \textit{s}/\textit{p}) is chosen. 
	In the simulations, we neglect explicit temperature dependence of the material properties. Therefore, Equation (\ref{eq:Kirchhoff}) is independent of $T$. This typically causes some deviations between theory and experiment \cite{song-2015}. 
	The resulting radiance at a given temperature, which can be measured in experiments, can thus be computed by weighing the emissivity by the Planckian radiance spectrum $R$ at the corresponding temperature, which can be expressed depending on the spectral variable as 
	\begin{equation}\label{eq:Planck-spectrum}
		R_{\nu}(\nu) = \frac{2h\nu^3/c^2}{e^{\frac{h\nu}{k_\textrm{B}T}} - 1} \quad\textrm{or}\quad R_{\bar{\nu}}(\bar{\nu}) = \frac{2h\bar{\nu}^3c^2}{e^{\frac{hc\bar{\nu}}{k_\textrm{B}T}} - 1}, 
	\end{equation}
	where $\bar{\nu}$ denotes wave number, defined as the inverse of the wavelength. The constants $h$, $c$, and $k_\textrm{B}$ denote the Planck constant, the speed of light, and the Boltzmann constant, respectively. 
	Kirchhoff's law of thermal radiation only holds (necessarily) true for reciprocal objects, however, the statement can be generalised \cite{guo-2022}. 

    \subsection{Computation of the T-matrices}\label{sec:methods_JCM}
	The T-matrices were computed using the FEM solver JCMsuite (JCM) \cite{pomplun-2007} for a range of frequencies. To this end, the three-dimensional meshes of the objects in question were constructed with an edge length at most as long as one fifth of the smallest considered wavelength in the exterior domain, and one fiftieth near sharp edges. Furthermore, the mesh had at least three vertices along each edge of the object. 
	The object was surrounded by a computational domain of a homogeneous, isotropic and lossless material, along with perfectly matched layer boundaries on all sides to mimic an open surrounding. 
	The response of the object to a set of incident multipolar fields was computed and subsequently expanded in outgoing vector spherical harmonics using the built-in \texttt{MultipoleExpansion} method. The expansion coefficients correspond to the T-matrix entries. 
	The number of incident multipolar fields and the expansion order of the outgoing fields (multipolar order) had to be chosen sufficiently large such that all relevant electromagnetic features could be encoded in the T-matrix. 
	More details can be found in the supplementary material. 
	The T-matrices resulting from iterating this approach over the selected wavelength range were stored together with metadata in the HDF5 data format \cite{asadova-2024} and are available via the T-matrix portal \cite{tmatrix-database}. 

    \subsection{Computation of metasurface emissivity}\label{sec:methods_treams}
	The python library \textsc{treams} was used to compute optical properties of a metasurface constructed from precomputed T-matrices \cite{beutel-2023-treams}. These T-matrices were converted to \texttt{PhysicsArray} objects and incorporated in a two-dimensional lattice, yielding an effective T-matrix constructed from the cluster T-matrix that represents the entire unit cell, allowing for multiple scatterers per unit cell. 
	The effective T-matrix of a unit cell in the lattice describes the interaction with all other unit cells \cite{zerulla-2022-homogenisation} and is used to obtain the S-matrix, which encodes the response of the metasurface to plane wave illuminations. 
	This procedure involves lattice sums \cite{beutel-2023-lattice} and a change of basis from the spherical to plane waves. The number of plane waves included in this basis have been limited based on their momentum component $k_{\sslash}$, parallel to the metasurface, to those contained in a spherical region in reciprocal space by the factor $b_{\max}$, given by $k_{\sslash,\max} \leq \frac{2\pi b_{\max}}{\Lambda}$. Here, $\Lambda$ denotes the largest of the lattice constants. This essentially limits the diffraction orders (including evanescent waves) taken into account. The parameter $b_{\max}$ should be chosen as large as possible to include as much of the information from the T-matrix in the resulting S-matrix. However, there is an upper bound to $b_{\max}$, depending on the multipolar order of the T-matrix. Exceeding this limit causes divergent behaviour (see supplementary material), like shown in the work by \textcite{egel-2017-truncation}, where the authors analysed the plane wave expansion of outgoing fields. 

	The S-matrix of the lattice, together with S-matrices representing other layers of a given metasurface, were combined to construct a single S-matrix representing the entire metasurface. S-matrices representing homogeneous layers, such as substrates, were constructed directly in \textsc{treams}. 
	From the combined S-matrix, the transmittance $\mathcal{T}$ and reflectance $\mathcal{R}$ upon plane wave illumination with frequency $\nu$, at polar angle $\theta$, azimuthal angle $\varphi$, and given polarisation $\tau$ were directly computed from the corresponding S-matrix entries. 
    The reflectance(transmittance) denotes the perpendicularly outgoing energy flux in the opposite(same) direction of the incident illumination, relative to the incident energy flux. The incident energy flux is given by the time average Poynting vector $\langle S_z\rangle$, where $S_z = \frac{1}{2}\textrm{Re}(\vec{E}^*\times\vec{H})\cdot\hat{z}$. The reflected and transmitted energy flux includes all scattered non-evanescent diffraction orders. 
    The directional, polarisation and frequency dependent absorptivity was consequently obtained as \begin{equation}\label{eq:abs}\mathcal{A}(\theta,\varphi,\tau,\nu) = 1 - \mathcal{R}(\theta,\varphi,\tau,\nu) - \mathcal{T}(\theta,\varphi,\tau,\nu). \end{equation}
    The emissivity was computed using Equations (\ref{eq:Kirchhoff},\ref{eq:abs}). \\*

    The thermal emissivities of two different metasurfaces, depicted schematically in Figure \ref{fig:metasurfaces_schematic}, were computed with the methods described above. 
	Section \ref{sec:Pt_slabs} discusses computations on the square lattice of platinum square plates on a silicon nitrite spacer layer and a tungsten substrate, which has been analysed in experiments by \textcite{costantini-2015}, to evaluate whether the methods yield physically accurate results. Simulations with the RCWA method were also performed for additional reference. 
	A second metasurface that has yet to be realised in an experiment was considered for evaluating the thermal radiation of a chiral system with the proposed method. The results are presented in Section \ref{sec:Ag_helices}. 

    \section{Square lattice of platinum square plates}\label{sec:Pt_slabs}
	The T-matrices of a platinum rectangular cuboid of 0.1~$\upmu$m in height and 0.855~$\upmu$m in length and width, surrounded by vacuum, were computed up to multipolar order 10 using the method described in Section \ref{sec:methods_JCM} for 121 frequencies linearly spaced in the interval from 45.3~THz to 143.2~THz (1512~cm\textsuperscript{-1} to 4776~cm\textsuperscript{-1}). The permittivity of platinum at these frequencies was taken from the Drude-Lorentz model computations by \textcite{rakic-1998}. By interpolating the T-matrix entries with a piecewise third order polynomial, the number of T-matrices was increased to 247 evenly spaced T-matrices across this frequency range. 
	
\begin{figure*}[t]
		\centering
		\makebox[\textwidth]{
			\includegraphics{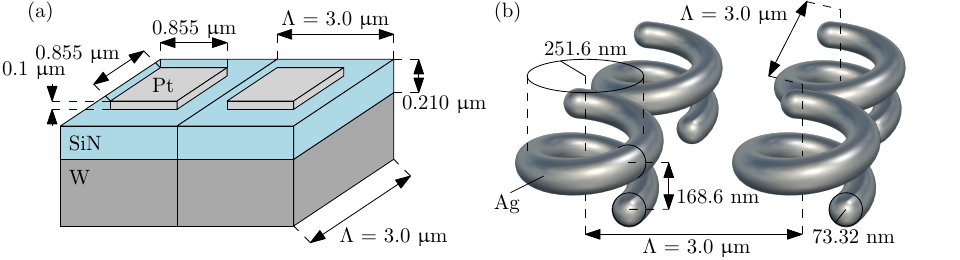}
		}
		\caption{(a) Schematic side view of the square lattice of platinum square plates, with materials and lengths indicated. 
			(b) Four unit cells of a square lattice of silver helices, exemplifying the considered chiral metasurface. 
		}
		\label{fig:metasurfaces_schematic}
	\end{figure*}

	The effective T-matrix was computed in the parity basis for a square lattice with a lattice constant of 3.0~$\upmu$m and converted to an S-matrix with the choice $b_{\max} = 3\sqrt{2}$. This S-matrix, along with the S-matrices of the homogeneous layers of the stratified system as shown in Figure \ref{fig:metasurfaces_schematic} were used to compute the absorption coefficients for each wavelength of the interpolated T-matrices, for 35 polar angles from 0 to 70$^\circ$ and azimuthal angles of 0, $\frac{\pi}{8}$, and $\frac{\pi}{4}$, for both \textit{s} and \textit{p} polarisations. The permittivities of the tungsten substrate and the silicon nitride spacer were taken from Drude-Lorentz model computations by \textcite{rakic-1998} and measurements on sputtered material by \textcite{kischkat-2012}, respectively. 
	
	\subsection{Results and discussion}
\begin{figure*}[t]
		\centering
		\makebox[\textwidth]{\includegraphics[width=.9\paperwidth]{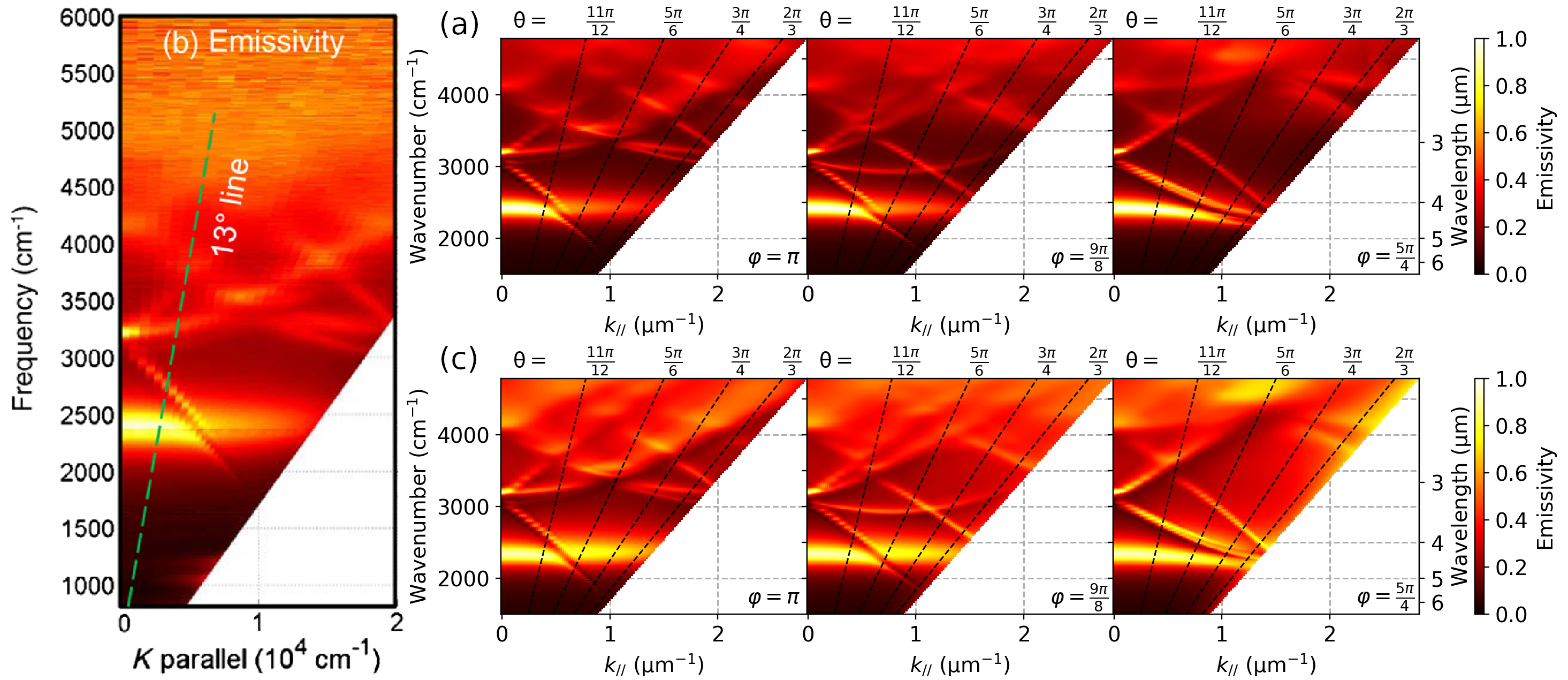}}
		\caption{
			(a) Polarisation averaged emissivities of the platinum square plate metasurface for azimuthal angles $\varphi\in\left\lbrace\pi,\frac{9\pi}{8},\frac{5\pi}{4}\right\rbrace$, computed using \textsc{treams} with T-matrices computed using JCM. 
            (b) Measured polarisation averaged emissivity data by \textcite{costantini-2015} (DOI: \texttt{10.1103/physrevapplied.4.014023}), measured at $\varphi = \pi$. 
			(c) Polarisation averaged emissivities of the platinum square plate metasurface computed using the RCWA method, for reference (see Section \ref{sec:performance_eval}). 
		}
		\label{fig:emissivity_Pt_metasurface}
	\end{figure*}

	The emissivity spectra from the computations on the platinum square plate metasurface are shown in Figure \ref{fig:emissivity_Pt_metasurface}. The angles of incidence have been converted to momentum components parallel to the surface, according to $k_{\sslash} = k_0\sin(\theta)$, in order to compare the results to those in the work by \textcite{costantini-2015}. 

    Both simulation results at azimuthal angle $\varphi = \pi$ show the same features as measured in the experiment. Most notable is the relatively broad emissivity feature with a maximum located just below 2500~cm\textsuperscript{-1}. 
	Its position and full width at half maximum can be found in Table \ref{tab:slab_metasurface_prop}, together with the relative radiance and the radiative efficiency in the ``useful window''. 
	The useful window has been defined as a 120~cm\textsuperscript{-1} wide spectral region around the emissivity maximum, with an opening angle of $13^\circ$ ($\Omega = 0.84$~sr) \cite{costantini-2015}. 
	The relative radiance is the emitted power in the useful window compared to a black body in the same window. The radiative efficiency is the emitted power in the useful window compared to the total emitted power of a black body. 
	
	\begin{table*}[t]
		\caption{Properties of the metasurface emissivity distribution, with measurement data and reported RCWA simulation taken from \textcite{costantini-2015}. }
		\makebox[\textwidth][c]{
			\begin{tabular}{l|cccc}
				& \parbox{2.6cm}{\centering Peak position\\(cm\textsuperscript{-1})} & \parbox{1.3cm}{\centering FWHM\\(cm\textsuperscript{-1})} & \parbox{3.2cm}{\centering Relative radiance\\in `useful window'} & \parbox{2.9cm}{\centering Radiative efficiency}\\
				&&&&\\[-1em]\noalign{\hrule height 1.6pt}
				Measurement & 2390 & 370 & 85 \% & 3.1 \%\\\hline
				Reported RCWA & 2350 & 250 & - & - \\\hline\hline
				\textsc{treams} & 2400 & 225 & 84.9 \% & 4.7 \%\\\hline
				RCWA & 2334 & 305 & 84.2 \% & 2.9 \%\\\hline\hline
				Black body & - & - &  100\% & 0.5 \%
			\end{tabular}
		}
		\label{tab:slab_metasurface_prop}
	\end{table*}

    From Table \ref{tab:slab_metasurface_prop}, it is evident that the emissivity peak positions obtained by simulation match the experimentally measured peak position rather well. The RCWA simulations deviate a bit more compared to the \textsc{treams} simulation, which is due to the convergence behaviour of the RCWA (see supplementary information). 
	All simulation methods underestimate the width of the peak. This can partly be explained by the fact that the permittivities used in simulation are valid at 298~K, whereas the measurement took place at an elevated temperature of 873~K. 
	The integrated quantities over the useful window are also reproduced quite well. The radiative efficiency of the metasurface being larger compared to that of the black body shows that the thermally radiated power of this metasurface is concentrated stronger in the useful window compared to a black body, which is desirable for applications. 
	Figure \ref{fig:emissivity_Pt_metasurface} also contains predicted emissivities for angles that were not measured, which show significant changes regarding the number and the width of emission lines. 
	
	The agreement between our proposed approach, the experimental results, and the RCWA simulations is a valuable validation of our methods and computer codes. 

    \subsection{Computational costs}\label{sec:performance_eval}
	Reference spectra using RCWA were computed to compare the accuracy and computational costs of the implementation of the T-matrix method in \textsc{treams} to an established method in computational photonics. RCWA was also used in the work by \textcite{costantini-2015}. 
	The RCWA reference spectra were computed using the \mbox{\textsc{fmmax}} library \cite{schubert-2023}. The structured layers were represented by a binary mask to model the permittivity contrast. The Fourier series coefficients were computed using this mask and the permittivities of the object and its surrounding material. 
	The fields were computed via the `pol'-method described in \textcite{liu-2012}, with applied damping of strong high frequency Fourier components. This ensured that the fields tangential to the metasurface could be determined with greater accuracy around permittivity transitions. 
	
	For further reference, a FEM simulation in the COMSOL software was performed at a fixed angle of incidence to compare the computational costs for a single wavelength. 
	
	All simulations were performed on the same hardware, featuring a $2\times$AMD EPYC 7453 processor, to enable a comparison between the computational costs of these methods. Total computation times (wall-clock time) and CPU-times, which reflect the speed-up from multiprocessing, were tracked. 
	
\begin{figure}[ht]
		\centering
		\includegraphics[width=\linewidth]{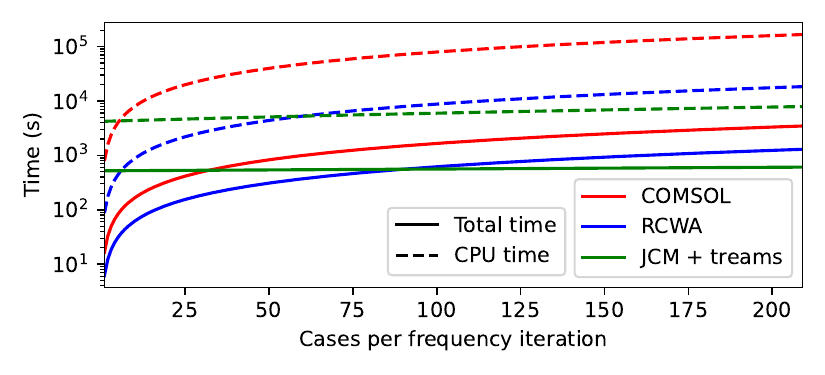}
		\caption{Computational costs per frequency iteration, as a function of the number of cases sampling the polar angle, azimuthal angle, and polarisation at each frequency. All methods scale approximately linearly with the number of cases, with FEM and RCWA having a much steeper slope than the \textsc{treams} simulation, which has a large initial offset due to the fixed computation time of the T-matrix. After about 61 cases per frequency, JCM + \textsc{treams} outperforms the other two methods in CPU-time. Fluctuations in the computation times are not visible on the logarithmic scale. }
		\label{fig:iteration_times}
	\end{figure}

	The costs of these computations can be appreciated in Figure \ref{fig:iteration_times} and Table \ref{tab:cost_per_freq}. 
	The method we propose has a fixed cost per frequency for computing the T-matrices through the FEM simulation. 
	After such cost has been paid, \textsc{treams} is comparatively very fast in sampling different incident angles and polarisations, but also layer thicknesses, lattice spacings and unit cell arrangements, for each frequency. 
	We denote one specific set of these parameters as `case' per frequency. 
	The computational time of \textsc{treams} increases linearly with the number of cases, but with a slope that is much smaller than those of RCWA or FEM. 
	This can be seen in Table \ref{tab:cost_per_freq} where the computation times per frequency per case for different methods are listed. Accordingly, Figure \ref{fig:iteration_times} shows crossings of the computational costs of JCM + \textsc{treams} with the other two methods, as the number of cases per frequency increases. 
	
	We highlight that such crossings are modest numbers of cases per frequency, that are likely to be insufficient for the characterisation of thermal radiation, in particular when working with very asymmetric metasurfaces \cite{overvig-2021} or systems that radiate in very narrow angular ranges \cite{greffet-2002}. 
	\begin{table}[ht]
		\centering
		\caption{Average computational cost for different methods, per frequency iteration, per case. }
		\begin{tabular}{l|cc}
			&&\\[-1em]
			Method & CPU-time (s) & Total time (s) \\\noalign{\hrule height 1.6pt}
			FEM (COMSOL) & $1.9\cdot10^2$ & 39 \\\hline 
			RCWA & 75 & 5.3 \\\hline
			\textsc{treams} & 0.42 & 0.42
		\end{tabular}
		\label{tab:cost_per_freq}
	\end{table}
	Additionally, using RCWA is rather awkward for some geometries, such as the helices investigated in the next section. Importantly, the number of cases per frequency, and hence the computational advantage of the proposed method, will quickly increase as more degrees of freedom are exploited for optimising the design, such as the lattice kind, lattice spacing, the material and the thickness of homogeneous layers embedding the lattice. 
	
	The memory used by the different methods was 14 GB, $\sim$2.4 GB, 19 GB and ~1 GB for any particular frequency iteration, for COMSOL, RCWA, JCM, and \textsc{treams}, respectively. 

    \section{Chiral metasurface}\label{sec:Ag_helices}
\begin{figure}[t]
		\centering
		\includegraphics[width=\linewidth]{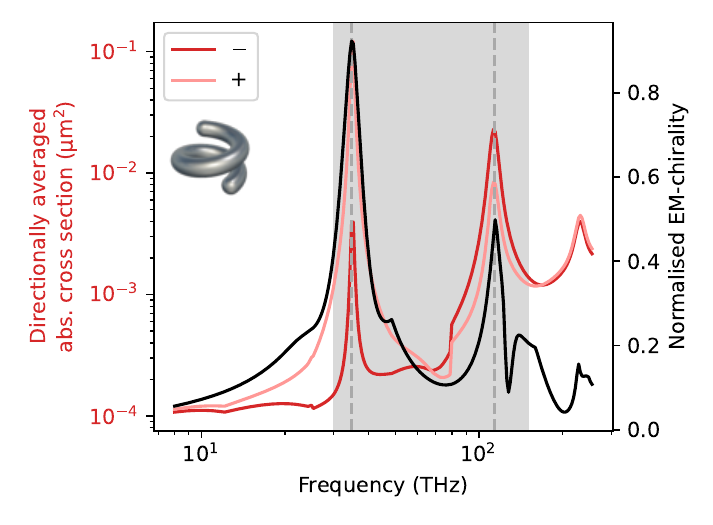}
		\caption{Frequency dependence of the helicity-dependent absorption cross section of a single silver helix, optimised for maximising its em-chirality. Different helicities are denoted by different shades of red, together with the frequency dependence of the normalised em-chirality of a single silver helix in black. 
		The grey shaded region indicates the frequency range used for lattice computations. The frequencies of the absorption maxima are indicated by the dashed vertical lines. }
		\label{fig:helix_characterisation}
	\end{figure}

	The chiral metasurface consisted of a square lattice of metallic silver helices in vacuum. 
	The parameters of the single helix were a radius of $r_h = 251.6$~nm and a wire radius of $r_w = 73.32$~nm with 1.407 right-handed turns at a pitch of $p = 168.6$~nm. The ends of the helix wire were rounded off by spherical caps. 
    These parameters were taken from a silver helix optimised for a large em-chirality \cite{fernandezcorbaton-2016}, adjusting for different ambient material parameters. The optimisation procedure was similar to the one by \textcite{garcia-santiago-2022}. 
	Selected properties of the isolated single helix are shown in Figure \ref{fig:helix_characterisation}. 
	This figure shows that the single helix has a strong chiral absorption around 34.8~THz, where the em-chirality reaches 92.3\% of its upper bound. An em-chirality of 100\% would correspond to an object that is invisible to one of the two helicities. Notably, the right-handed helix interacts predominantly with left-handed helicity (+) light at the strong chiral absorption. 
	Another absorption peak is located at 114~THz, favoured by light with the right-handed helicity (-), albeit with a smaller difference in the absorption between the two helicities, resulting in a lower em-chirality. 
	
    The T-matrices of the helices contained 5 multipolar orders and were passivated over the considered frequency range by imposing positive semidefiniteness of the imaginary unit times the skew-Hermitian part of the corresponding reactance matrix \cite{ru-2013}. This corresponds to setting the optical gain to zero. 
    This was necessary because of numerical noise in the T-matrix from the more complicated geometry and the relatively small size of the helix compared to the wavelength. 
	As described above, the effective T-matrix of the lattice was computed using \textsc{treams} and converted to an S-matrix. The lattice constant was set to 3~$\upmu$m and the number of diffraction orders was limited by $b_{\max} = 3$. 
	The reflection and transmission coefficients of the lattice were computed for 12 polar angles from 0 to $\pi$, azimuthal angles of 0, $\frac{\pi}{8}$, $\frac{\pi}{4}$, and $\frac{3\pi}{8}$, for both left- and right-handed circular polarisations. 
	The emissivity was computed from the absorptivity. To analyse the circular polarised light, the computations were in this case performed in the helicity basis. 

    \subsection{Results and discussion}
\begin{figure*}[t]
		\centering
		\includegraphics[width=.9\linewidth]{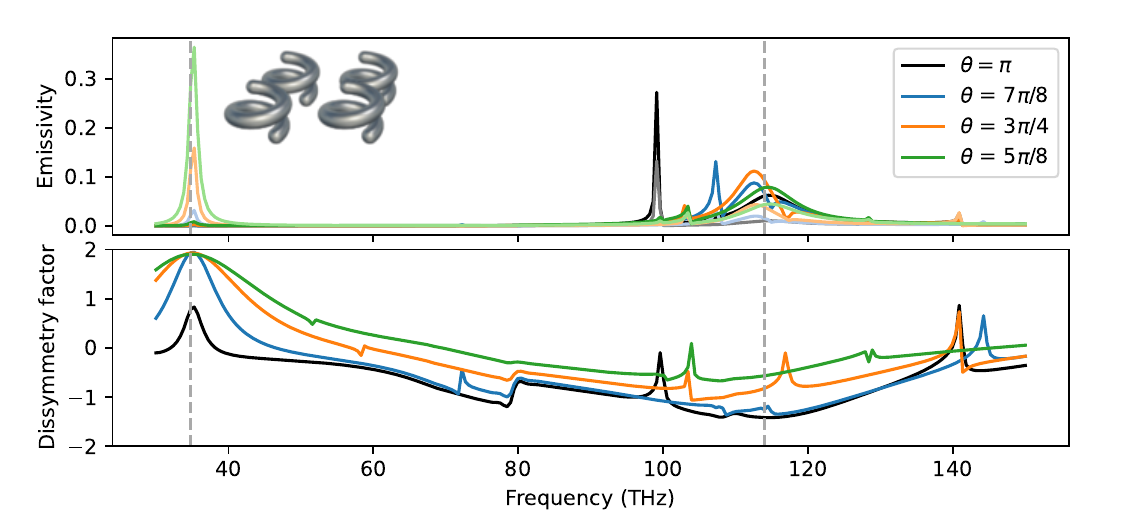}
		\caption{Helicity-dependent emissivity of light from the two-dimensional array of silver helices and dissymmetry factor for different polar angles, in the frequency range marked by the shaded region from Figure \ref{fig:helix_characterisation}. Right-handed and left-handed circular polarisations are indicated by solid and pale colours, respectively. }
		\label{fig:emissivity_chiral_metasurface}
	\end{figure*}

	To compare the emissivity of left- and right-handed circular polarised light, the normalised emissivity difference, also referred to as the `thermal g-factor', 
	\begin{equation}
		g_{\textrm{th}} = 2\frac{\mathcal{E}_+ - \mathcal{E}_-}{\mathcal{E}_+ + \mathcal{E}_-}
	\end{equation}
	was computed. The emissivities $\mathcal{E}_\pm$ together with $g_{\textrm{th}}$ are shown in Figure \ref{fig:emissivity_chiral_metasurface}. 
	This figure shows that the emissivity peaks in frequency regions where the single helix has a large absorption cross section in Figure \ref{fig:helix_characterisation}. 
	At 35.3~THz, $g_\textrm{th}$ reaches its maximum of almost 2, resulting in almost helicity-pure thermal radiation at this frequency. 
	The emissivity maximum corresponding to the 34.8~THz absorption cross section peak is slightly blueshifted, while the emissivity maximum corresponding to the 114~THz absorption cross section peak is redshifted compared to the single helix. 
	It is remarkable that the emissivity, in combination with the large dissymmetry factor, results in thermal radiation with strongly favoured and opposite helicity at these frequencies. 

    Because of the chosen lattice spacing of 3~$\upmu$m, the spectrum also shows several lattice resonance peaks, depending on the angle of incidence, at which the emissivity is increased noticeably compared to nearby frequencies. 
	The dissymmetry factor is also affected by the lattice resonances, showing deviations from the otherwise smooth behaviour. 
	The lattice resonance with the smallest transverse momentum, corresponding to normal incidence, lies at 99.9~THz. 
	For incidence at an angle, the lattice resonances are weaker and lie at lower frequencies. The feature around 80~THz in the dissymmetry factor plot is due to a numerical artefact from concatenating two sets of T-matrices to a single frequency range. 

    Overall, emissivity of negative helicity light is stronger at large frequencies, while the emissivity of positive helicity light is stronger at smaller frequencies, as can be seen in the bottom row of Figure \ref{fig:emissivity_chiral_metasurface}. 

    \section{Conclusion and outlook}\label{sec:conclusion}
	In this work, we analysed thermal radiation from metasurfaces by using a T-matrix-based method in combination with the directional Kirchhoff law. We showed that the implementation of the T-matrix method in \textsc{treams}, with T-matrices computed using a FEM solver, is able to reproduce experimental measurements. 
	The more critical points of the method are the computation of the T-matrix, which has to be of sufficient multipolar order to achieve accurate results, and the convergence of the lattice sums, which requires one to limit the momentum transfer in lattice interactions. The method is expected to be inadequate when the objects in the planar lattice invade each other's smallest circumscribing sphere. 
	
	The method is more efficient than established methods based on RCWA or FEM as soon as the number of cases per frequency exceeds some threshold. For a given metasurface, the number of cases is the total number of radiation directions and polarisations for which the emissivity is computed. 
	Beyond that threshold, the fixed cost of computing the T-matrices already pays off, since for a fixed frequency, each case only requires a computation in \textsc{treams}, which is comparatively very fast. For the metasurface measured in \cite{costantini-2015}, the threshold was 61 cases for RCWA and 6 cases for FEM. We expect that the threshold would be exceeded for most metasurfaces. 
	Moreover, the optimisation of metasurfaces for the control of thermal radiation may require to evaluate different kinds of lattices, a range of lattice constants and/or variations of the thickness and material of homogeneous layers. Exploiting these degrees of freedom will significantly increase the number of cases, making the implementation in \textsc{treams} a useful and competitive tool for designing metasurfaces for thermal photonics. 
	The method could also be applied to metasurfaces whose unit cells are made of atomic clusters, such as molecular crystals or arrays of graphene patches, for example. For such cases, the T-matrix can be obtained using quantum-chemical methods. 

    \section{Acknowledgements}
	I.F.C. and C.R. acknowledge support by the Helmholtz Association via the Helmholtz program ``Materials Systems Engineering'' (MSE). 
	L.R. acknowledges support by the Karlsruhe School of Optics \& Photonics (KSOP). 
	L.R., I.F.C., and C.R. acknowledge support by the Deutsche Forschungsgemeinschaft (DFG, German Research Foundation) -- Project-ID 258734477 -- SFB 1173. 
	M.N. and C.R. acknowledge support by the KIT through the ``Virtual Materials Design'' (VIRTMAT) project. 
	The authors are grateful to the company JCMwave for their free provision of the FEM Maxwell solver JCMsuite. 
	
	\bibliography{bibfile}
    
\end{document}